\documentclass[3p,twocolumn]{elsarticle}

\usepackage{hyperref}

\usepackage[symbol]{footmisc}

\RequirePackage{graphicx}
\RequirePackage[caption=false]{subfig}
\RequirePackage{xcolor}

\usepackage{bm}
\usepackage{xspace}
\usepackage{multirow}
\usepackage{xcolor}
\usepackage[british]{babel}
\usepackage{hyphenat}
\usepackage{amsmath,amssymb}
\usepackage{physics}
\usepackage{comment}
\usepackage{array}

\usepackage{graphicx} 
\usepackage{caption}
\usepackage{subfig}
\usepackage{float} 
\graphicspath{{figures/}} 
\usepackage{booktabs}
\usepackage{hyperref}

\usepackage{eurosym}

\newcommand{\beq}{\begin{equation}}
\newcommand{\eeq}{\end{equation}}

\def\sib{{\textsc{Sibyll}\,2.3d}\xspace}
\def\qgs{\textsc{QGSJet}-II.04\xspace}
\def\epos{\textsc{EPOS-LHC}\,\xspace}
\newcommand{\conex}{\textsc{Conex}}
\newcommand{\fluka}{\textsc{Fluka}}

\def\Nmu{$N_\mu$\xspace}%
\def\LambdaMu{$\Lambda_\mu$\xspace}
\def\Xmax{$X_{\rm max}$\xspace}

\newcolumntype{C}[1]{>{\centering\arraybackslash}m{#1}}

\journal{Physics Letters B}

\begin{document}

\begin{frontmatter}

\title{Proton-air interactions at ultra-high energies in muon-depleted air showers with different depths}

\author[IGFAE]{L. Cazon}
\author[LIP,IST]{R. Concei\c{c}\~ao}
\author[IGFAE]{M. A. Martins \corref{mam}}
\ead{miguelalexandre.jesusdasilva@usc.es}
\author[IGFAE]{F. Riehn}

\cortext[mam]{Corresponding author}

\address[IGFAE]{Instituto Galego de F\'isica de Altas Enerx\'ias (IGFAE), University of Santiago de Compostela, Rúa de Xoaquín Díaz de Rábago, Santiago de Compostela, Spain}
\address[LIP]{Laboratório de Instrumentação e Física Experimental de Partículas (LIP), Lisbon, Portugal}
\address[IST]{Instituto Superior T\'{e}cnico (IST), Universidade de Lisboa, Lisbon, Portugal}

\begin{abstract}

The hardness of the energy spectrum of neutral pions produced in proton-air interactions at ultra-high energies, above $10^{18}$ eV, is constrained by the steepness of the shower-to-shower distribution of the number of muons in muon-depleted extensive air showers.
\par
In this work, we find that this steepness, quantified by the parameter \LambdaMu, evolves with the depth of the shower maximum, $X_{\max}$, assuming a universal value for shallow showers and an enhanced dependence on the high-energy hadronic interaction model for deep showers. We show that $X_{\max}$ probes the so-called hadronic activity of the first proton-air interaction, thus allowing direct access to the energy spectrum of neutral pions in different regions of its kinematic phase space.
\par
We verify that the unbiased measurement of \LambdaMu is possible for realistic mass composition expectations. Finally, we infer that the statistical precision in \LambdaMu required to distinguish between hadronic interaction models can be achieved in current extensive air shower detectors, given their resolution and exposure.
\end{abstract}

\end{frontmatter}


\section{Introduction}
\label{sec:intro}
The interactions between ultra-high-energy cosmic rays (UHECRs) and the nuclei in the Earth's atmosphere can occur at ten times the center-of-mass energies attained at the LHC. As such, those interactions have been promoted as a unique opportunity to make measurements relevant to particle physics beyond the reach of human-made accelerators. However, the measurement of the proton-air cross-section remains the only one performed to date \cite{ThePierreAuger:2012prl}.
\par
Recently, it was shown that the fluctuations in the muon content of extensive air showers (EAS) are sensitive to fluctuations in the energy sharing among secondaries of the first $p$-air interaction~\cite{Cazon:2018gww}. Additionally, the distribution of the number of muons in proton-induced EAS has a so-called \textit{tail} towards low values, constituted by muon-depleted showers. The steepness of this tail, measured by $\Lambda_\mu$, probes the production cross-section of neutral pions in the forward region of the kinematic phase-space of proton-air interactions~\cite{Cazon:2020jla}. This measurement is resilient to different composition scenarios.
\par
In this work, we argue that the evolution of $\Lambda_\mu$ with the depth of the shower maximum, \Xmax, probes the so-called {\it hadronic activity} of the first $p$-air interaction at center-of-mass energies above 100 TeV. In this context, {\it hadronic activity} encompasses the multiplicity of hadronically interacting particles, the fraction of energy in the hadronic channel and inelasticity of the $p$-air interaction. 
For the first time, we demonstrate that the energy spectrum of neutral pions can be constrained in different regions of the kinematic phase space of $p$-air interactions, which are not covered in accelerator experiments.

\section{Evolution of \LambdaMu with \Xmax}\label{sect:lambda_mu}
\begin{figure}[!h]
\centering
\includegraphics[width = \columnwidth]{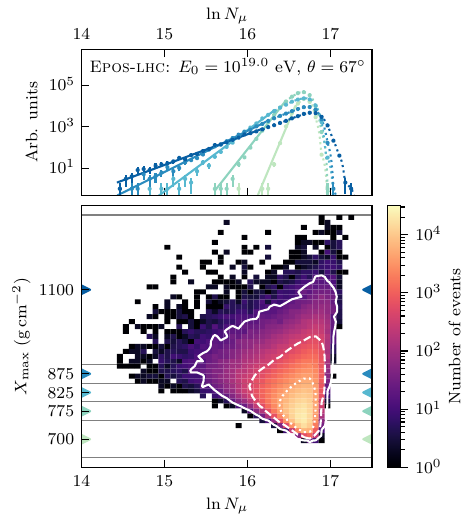}
\caption{Lower panel: Joint distribution of (\Xmax, $\ln N_\mu$), $f$(\Xmax, $\ln N_\mu$). The white contours contain $68 \%$ (dotted), $95 \%$ (dashed) and $99.7\%$ (solid) of the shower simulations. The horizontal lines represent bins in \Xmax with edges $\{650, \,750, \,800, \,850, \,900, \,1300\} \mathrm{\,g\,cm^{-2}}$ and centers indicated by the triangular markers. Higher values of \Xmax are represented by darker shades of blue and lower ones by light shades of green. Upper panel: distribution of $\ln N_\mu$ for each bin in \Xmax, $f_X( \ln N_\mu)$, matching the colour of the corresponding \Xmax bin. Details of the simulations are given in the main text.}
\label{fig:lnnmu_Xmax_joint_dist_proton_epos}
\end{figure}

In proton-induced extensive air showers, the fluctuations of the depth of the shower maximum, $X_{\max}$, and those of the number of muons reaching the ground level, $N_\mu$, are coupled, as both observables are sensitive to the multiparticle production properties of high-energy hadronic interactions. This coupling is clear from the joint distribution function $f$(\Xmax, $\ln N_\mu$), shown in the lower panel of Figure~\ref{fig:lnnmu_Xmax_joint_dist_proton_epos}. This figure was produced using $10^6$ proton-induced showers simulated with \conex{} v7.50~\cite{Bergmann:2007conex, Pierog:2004conex} with primary energy $E_0 = 10^{19}\,$eV and zenith angle $\theta = 67^\circ$, using the high-energy hadronic interaction model \epos{}~\cite{Pierog:2013ria}. Inclined showers were chosen to mitigate artificial correlations between \Xmax and $\ln N_\mu$ through the truncation of the shower development.
\footnote{Furthermore, particles above $E_{\text{th}} = 0.005 \times E_0$ and their interactions are tracked individually, while below it, the longitudinal shower profile numerically solves cascade equations. The ground level was set to $1\,400\,$m a.s.l, the average height of the Pierre Auger Observatory~\cite{ThePierreAuger:2015rma}, corresponding to an average vertical depth of $X_{\text{gr}} = 880\,\mathrm{g\,cm^{-2}}$. The number of muons at the ground level, $N_\mu$, is defined by the value of the muon longitudinal profile at a slanted depth of $X = X_{\text{gr}} \sec \theta$. Only muons with $E_{\mu}> 1\,$GeV were considered. The value of \Xmax is taken from a Gaisser-Hillas fit to the longitudinal profile of all charged particles.}
\par
Preliminary studies have partially exploited $f$(\Xmax, $\ln N_\mu$) to constrain the macroscopic multi-particle production properties of the highest energy hadronic-interactions~\cite{Goos:2023opq}. Although this macroscopic picture is useful, it does not allow the probing of more fundamental quantities, such as the cross-section for the production of neutral pions, which certainly impacts the development of the electromagnetic cascade through their decay into photons. Furthermore, the correlation between $X_{\max}$ and $N_\mu$ is decreased by the fluctuations of the depth of the $p$-air interaction, $X_0$, which are independent of muon production. However, since $X_0$ cannot be experimentally accessed, we carry our investigations using \Xmax directly.
\par
To examine the evolution of the distribution of $\ln N_\mu$ with $X_{\max}$, we bin the ensemble of shower simulations in the ranges of \Xmax discriminated in the lower panel of Figure~\ref{fig:lnnmu_Xmax_joint_dist_proton_epos}. Each bin corresponds to a colour code, ranging from light green (shallow \Xmax) to dark blue (deep \Xmax). We denote the distribution of $\ln N_\mu$ in a given \Xmax bin by $f_X(\ln N_\mu) = \int_{X-\Delta}^{X+\Delta} f(\ln N_\mu,X_{\max}) \dd{X_{\max}}$, where $X$ is the center of the bin and $2\Delta$ the width. The distributions $f_X(\ln N_\mu)$ are plotted in the upper panel of Figure~\ref{fig:lnnmu_Xmax_joint_dist_proton_epos}, matching the colour of the corresponding \Xmax bin. Moreover, we performed an unbinned likelihood fit to the left tail of $f_X(\ln N_\mu)$ using the function $y = A_\mu \exp \left\{ \ln N_\mu / \Lambda_\mu \right\}$. The tail is defined by all values of $\ln N_\mu$ below the $10 \%$ quantile of $f_X(\ln N_\mu)$ and it is represented by the coloured solid lines in the upper panel of Figure~\ref{fig:lnnmu_Xmax_joint_dist_proton_epos}. It is clear that the deeper the shower, the flatter the tail of the distribution of $\ln N_\mu$ towards low values. We checked that this trend is not due to effects related to muon flux attenuation in the atmosphere, by verifying that the trend is equally present when one analyses the distribution of the number of muons at production, at the shower axis~\cite{Andriga:2012_muonprofile, Cazon:2023_zombie}. Likewise, this behaviour is qualitatively independent of the binning in \Xmax. The chosen bin edges: $\{650, 750, 800, 850, 900, 1300\}\,\mathrm{g\,cm^{-2}}$ are independent of the hadronic interaction model and of the primary composition, hence independent of the detailed shape of the distribution of \Xmax, while allowing a precise characterisation of the features of the distribution of $\ln N_\mu$. Additionally, their widths are uneven to accommodate a similar number of events in each bin in $X_{\max}$ and of the order of a typical resolution attained in a measurement of $X_{\max}$~\cite{ThePierreAuger:2014prd}.
\par
\begin{figure}[!h]
\centering
\includegraphics[width = \columnwidth]{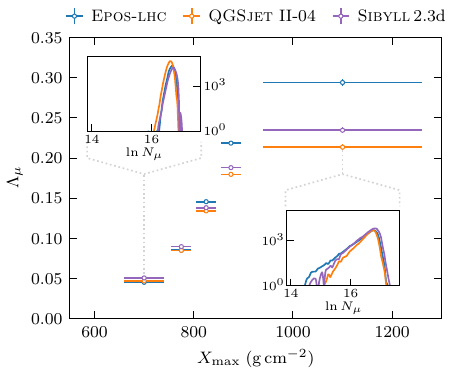}
\caption{ \LambdaMu as a function of \Xmax for the hadronic interaction models \epos, \qgs{} and \sib{}, for proton-induced EAS with $E_0 = 10^{19}$ eV and $\theta = 67^\circ$. The distributions of $\ln N_\mu$ for the shallowest and deepest bins in \Xmax are shown in the left and right insets, respectively.}
\label{fig:Lambda_mu_per_Xmax_proton_perfect_resolution}
\end{figure}

The evolution of the steepness of the tail of the distribution of $\ln N_\mu$ with \Xmax is quantified through the change in the parameter $\Lambda_\mu$, defined in the previous paragraph. Note that smaller (larger) values of \LambdaMu correspond to a steeper (flatter) tail. We have verified that the value of $\Lambda_\mu$ is independent of the choice of the quantile defining the tail provided it is below the bulk of the distribution. The values of \LambdaMu in each bin of \Xmax are shown in Figure~\ref{fig:Lambda_mu_per_Xmax_proton_perfect_resolution} for three different high-energy hadronic interaction models, \epos{} (blue), \qgs{}~\cite{qgs} (orange) and \sib{}~\cite{sib23d} (purple). The insets of the same figure display the distributions of $\ln N_\mu$ for the shallowest (left inset) and deepest (right inset) bins of \Xmax.
\par
\par
Evidently, there is a monotonic increase of $\Lambda_\mu$ with \Xmax, irrespective of the high-energy hadronic interaction model. Furthermore, the predicted value of $\Lambda_\mu$ is mildly model-independent for shallower showers and strongly model-dependent in deeper showers, where the relative model difference reaches $30 \%$. For reference, there is a model dependence of $20\%$ in \LambdaMu if extracted by fitting the distribution of the number of muons integrated over all \Xmax. The described behaviours are qualitatively independent of the widths of the bins in $X_{\max}$, provided the same region of the $X_{\max}$ distribution is being probed.
\par
The evolution of $\Lambda_\mu$ with \Xmax was further validated with $1\,000$ full Monte Carlo CORSIKA~\cite{CORSIKA} simulations and shown to be independent of the energy threshold of the muons in the simulation\footnote{Low energy hadronic interactions were handled with \fluka~\cite{fluka,fluka2}.}. Moreover, we verified that $\Lambda_\mu$ has a $15 \%$ dependence on the zenith angle $\theta$ for $\theta > 45^\circ$. This lower bound on $\theta$ ensures the full development of the muonic cascade before reaching the ground. Nevertheless, the qualitative evolution of $\Lambda_\mu$ with \Xmax and that of its model dependence are independent of $\theta$. Finally, we verified that the dependence of $\Lambda_\mu$ on the primary energy \footnote{To keep the \Xmax binning, showers with different primary energies were binned in $X_{19} = X_{\max} - 58 \log_{10}( E_0 / 10^{19} \,\mathrm{eV} )$, to correct for the elongation rate.} can be monotonically parameterised, ensuring the validity of the results presented at fixed energy.



\section{Connection with the first interaction}
\label{sect:1st-int}
The connection between \LambdaMu and the hardness of the spectrum of neutral pions produced in the first $p$-air interaction was established in~\cite{Cazon:2020jla}. In particular, first $p$-air interactions in which secondary neutral pions carry a significant fraction of the primary energy give rise to hadronic showers with a smaller muon yield as there is less energy available in the hadronic component for muon production. Hence, a harder energy spectrum of neutral pions leads to an increased probability of showers with low $N_\mu$, thus flattening the tail of the distribution of the number of muons.

\par
\begin{figure*}[!t]
\centering
\includegraphics[width= \linewidth]{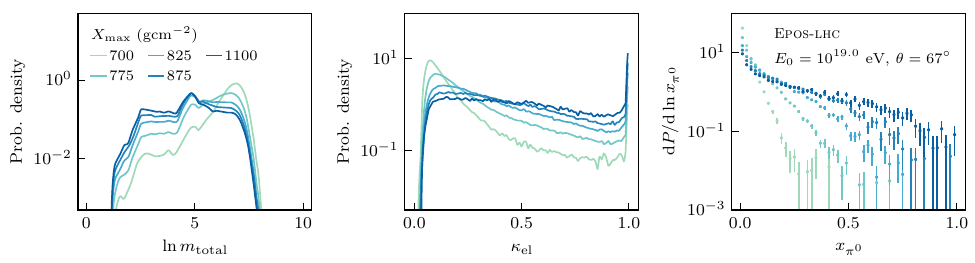}
\caption{Distributions of the total multiplicity, $\ln m_{\text{total}}$ (left panel), elasticity, $\kappa_{\text{el}}$ (middle panel) and the energy spectrum of neutral pions of the first $p$-air interaction (right panel), per bin in \Xmax. Details of the simulations are given in the text.} \label{fig:mean_1stInt_per_xmax_logNmu_bin_epos}
\end{figure*}

\par
The connection between the moments of the distributions of \Xmax and $\ln N_\mu$ and those of the distributions of macroscopic quantities of the first interaction, such as the multiplicity, elasticity and other derived quantities have been investigated~\cite{Goos:2023opq,Matthews:2005sd,Ulrich:2010rg}. To understand the evolution of $\Lambda_\mu$ with $X_{\max}$, we look at the distributions of total multiplicity, $m_{\text{total}}$, elasticity, $\kappa_{\text{el}}$ and the energy spectrum of neutral pions of the first interaction, in each bin of $X_{\max}$, as depicted in Figure~\ref{fig:mean_1stInt_per_xmax_logNmu_bin_epos}. In the laboratory frame, the fraction of energy carried by each neutral pion of the first interaction is denoted by $x_{\pi^0}$.
\par
It is clear that deeper showers tend to correspond to more elastic first interactions with lower multiplicities, where energy is more asymmetrically distributed among secondaries. These interactions are thus characterised by lower hadronic activity. The reciprocal is also true: shallower showers tend to correspond to more deeply inelastic first interactions with higher multiplicity, thus higher hadronic activity. The observation that higher multiplicities produce showers that develop faster has been explored before in ~\cite{Ulrich:2010rg}. Likewise, increasing the elasticity of the highest energy $p$-air interactions leads to showers that develop more slowly, hence with deeper \Xmax values. Therefore, selecting showers by \Xmax permits a continuous probing of the \emph{hadronic activity} of the first $p$-air interaction, despite the uncorrelated fluctuations of its depth in the atmosphere. Note that the details of the causal connection between hadronic activity and the depth of the shower, on a shower-to-shower basis, need to be explored using shower simulations or deep learning models as explored in~\cite{Goos:2023opq}.
\par
Lastly, from the right panel of Figure \ref{fig:mean_1stInt_per_xmax_logNmu_bin_epos}, it is evident that deeper showers tend to correspond to first interactions with fast neutral pions. These immediately decay into high-energy photons that dominate the electromagnetic profile and yield deeper showers. So, the energy spectrum of neutral pions constrained to deep showers is harder, flattening the tail of the distribution of $\ln N_\mu$. Therefore, the production spectrum of neutral pions can be constrained, as a function of the hadronic activity of the first $p$-air interaction, through the evolution of $\Lambda_\mu$ with \Xmax, in phase-space regions unconstrained by accelerator data. Note, however, that the divergence of the value of $\Lambda_\mu$ obtained with \epos{}, relative to other models, in deeper bins in $X_{\max}$ cannot be solely explained from the differences in the spectrum of neutral pions of the first proton-air interaction.
\par
Finally, we propose a possible explanation for the increasing model dependence of \LambdaMu with \Xmax. Shallow showers are characterized by high hadronic activity (high multiplicity of secondaries among which energy is more evenly distributed), so secondaries are produced in kinematic
regions more accessible to accelerator experiments. Thus the shape of the energy spectrum of secondary neutral pions is better constrained by data, yielding similar values of \LambdaMu across models.
In contrast, deep showers have reduced hadronic activity (low
multiplicity of secondaries among which energy is very asymmetrically distributed). Thus,
leading particles tend to occupy the forward region of the p-air
interaction's kinematic phase space, where the models are less constrained by accelerator data. Due to this freedom, the shape of the energy spectrum of neutral pions becomes more dependent on the particular physical mechanisms employed by each model, resulting in an enhanced model dependence of \LambdaMu.

\section{Experimental feasibility of the measurement of $\Lambda_\mu$ as a function of \Xmax}
The ability to distinguish the predictions of the different high-energy hadronic interaction models for $\Lambda_\mu$ depends on at least three aspects: 1) on the composition of the flux of cosmic rays (and its evolution with the primary energy) since the proton fraction determines the exponential tail of the distribution of $\ln N_\mu$; 2) on the resolution in the reconstruction of the shower observables: $\ln N_\mu$, $E_0$ and $X_{\max}$, which partially de-correlates \Xmax and $N_\mu$, widens the distribution of $\ln N_\mu$ and causes migrations between the \Xmax bins, and 3) on the number of events in the full ensemble of measured showers.
\par
The feasibility of the measurement of \LambdaMu in each bin of \Xmax was quantified for two different composition scenarios. For primary energies in $[10^{18}, 10^{18.5}]$ eV, these scenarios correspond to the ones with the lowest and highest proton fractions given by the 4-mass fraction fit to Auger data presented in \cite{Tkachenko:2023icrc}. Each scenario is mimicked by sampling from $10^5$ proton, helium, nitrogen and iron-induced simulated showers, in proportions: p : He : N : Fe $= 7:1:2:0$ and $1:3:1:0$. Furthermore, Gaussian resolutions with $\sigma (N_\mu) / N_\mu = 20 \%$ \cite{ThePierreAuger:2015prd} and $\sigma(X_{\max}) = 20 \mathrm{\,g\,cm^{-2}}$ \cite{ThePierreAuger:2014prd} were considered in the reconstruction of $\ln N_\mu$ and \Xmax, respectively. In a real scenario, the composition interpretation of the cosmic ray flux depends on the hadronic interaction model and evolves with the primary energy \cite{ThePierreAuger:2017prd}. Moreover, the falling cosmic ray spectrum in interplay with the composition couples the composition evolution with the precision of the measurement of \LambdaMu via the limiting statistics. For simplicity, we decouple these effects by studying extreme composition cases in a small bin of primary energies.
\par
\begin{figure}[!h]
\centering
\includegraphics[width = \columnwidth]{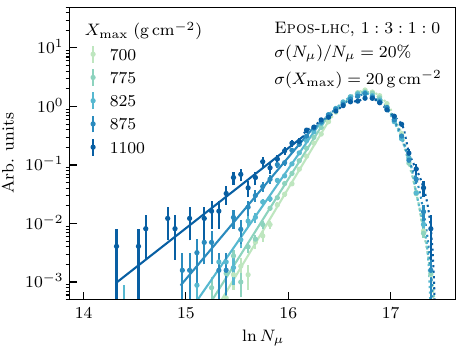}
\caption{Fitted distributions of $\ln N_\mu$ for each \Xmax bin for a mixture of p : He : N : Fe in proportion 1 : 3 : 1 : 0, assuming Gaussian resolutions of $20 \%$ in $N_\mu$ and $20 \,\mathrm{g\,cm^{-2}}$ in $X_{\max}$. The $10^5$ \conex{} showers, for each primary, were simulated with a primary energy of $E_0 = 10^{19}\,$eV and $\theta = 67^\circ$ using the high-energy hadronic interaction model \epos{}.} \label{fig:lnNmu_per_Xmax_mixed_comp_wresol_heatmap_epos}
\end{figure}
\par
Figure~\ref{fig:lnNmu_per_Xmax_mixed_comp_wresol_heatmap_epos} displays the distribution of $\ln N_\mu$ in each bin in $X_{\max}$ for the composition scenario $1:3:1:0$ and the aforementioned resolutions. The tails of each distribution $f_X(\ln N_\mu)$ were fitted as described in Section~\ref{sect:lambda_mu}. Despite the smearing introduced by the resolution in $f_X(\ln N_\mu)$, the tails due to proton primaries remain visible and evolve with \Xmax in agreement with the results presented in Section~\ref{sect:lambda_mu}.
%
\par
\begin{figure}[!t]
\centering
\includegraphics[width = \columnwidth]{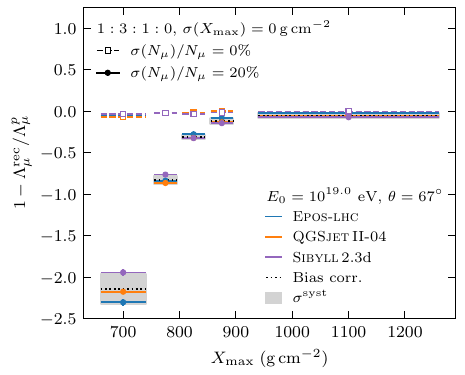}
\caption{Relative bias in $\Lambda_\mu$ as a function of the \Xmax bin, for mixed primary composition in proportions $1:3:1:0$, for a perfect (dashed) and $20 \%$ (solid) resolutions in the reconstruction of $N_\mu$. The average bias correction is represented by the dotted black line. The grey band represents the systematic uncertainty on the hadronic interaction model.} \label{fig:LambdaMu_per_Xmax_mixed_comp_residualsToProton}
\end{figure}

We quantify the impacts of the reconstruction resolution and the presence of heavier primaries on \LambdaMu separately. That is, for the composition scenario $1:3:1:0$, we fit the tail of
each distribution of $\ln N_{\mu}$ to extract the reconstructed value of $\Lambda_{\mu}$, $\Lambda_{\mu}^{\text{rec}}$, for two cases: a perfect reconstruction resolution in $N_\mu$ and $\sigma(N_\mu) / N_\mu = 20 \%$. The resolution in \Xmax can be regarded as an additional smearing in $N_\mu$, so here we set it to $\sigma(X_{\max}) = 0\,\mathrm{g\,cm^{-2}}$. For these two cases, we compute the residuals of $\Lambda_{\mu}^{\text{rec}}$ relative to the value of $\Lambda_\mu$ obtained for proton-induced showers, henceforth $\Lambda_{\mu}^p$, as shown in Figure~\ref{fig:LambdaMu_per_Xmax_mixed_comp_residualsToProton}, as a function of \Xmax, using the three high-energy hadronic interaction model \epos, \qgs{} and \sib{}. Dashed lines correspond to a mixed composition with perfect resolution, and the solid lines to an assumed resolution of $20 \%$ in the reconstruction of $N_\mu$. The black dotted line represents the model-averaged bias, and the grey line represents the difference between high-energy hadronic interaction models. We take this dependence as a systematic uncertainty.
\par
The bias induced by a mixture of primary masses on \LambdaMu is less than $5 \%$, irrespective of the \Xmax bin. The small bias can be explained by the superposition principle: for a primary with a mass number $A$, the distribution of the number of muons is approximately the $A$-fold self-convolution of the one present for proton showers, thus suppressing the distribution's tail. Hence, the tail of the distribution of $\ln N_\mu$ for all showers is dominated by the proton-induced ones, ensuring the applicability of the results of Section~\ref{sect:lambda_mu} to realistic compositions. On the other hand, the convolution of the detector resolution with the distribution of $\ln N_\mu$ greatly biases $\Lambda_\mu$ in shallower showers since the tail of the distribution is very steep. Additionally, for the shallow showers, the bias in \LambdaMu is model-dependent. For deeper showers, the bias in \LambdaMu reduces to about $10 \%$, allowing for an accurate estimation of \LambdaMu. Detailed strategies to de-convolve the detector resolution from the physical fluctuations of $N_\mu$ could mitigate the bias in \LambdaMu for shallower showers. However, for the sake of simplicity, we apply an average bias correction to \LambdaMu and take the model dependence as a systematic uncertainty. This uncertainty in \LambdaMu is greater than the difference between hadronic interaction models for showers with $X_{\max} < 800 \,\mathrm{g\,cm^{-2}}$, thus preventing model discrimination. Nevertheless, a measurement of \LambdaMu for shallow showers could still allow the simultaneous exclusion of all models. Finally, these observations also apply to the scenario with the higher fraction of proton primaries.
\par
\par
For showers with $X_{\max} > 800\,\mathrm{g\,cm^{-2}}$, we determined the minimum number of events, $n_{\min}$, necessary to distinguish between $\Lambda_\mu$ predictions from different hadronic interaction models. The distinction threshold was set to $1$ and $3\,\sigma$, corresponding to $n_{\min}^{1 \sigma}$ and $n_{\min}^{3 \sigma}$ events, respectively. The error in \LambdaMu was computed via a bootstrapping method to ensure its accurate estimation for low statistics. The values of $n_{\min}$ for the proton-rich and proton-depleted composition scenarios can be found in Table~\ref{tab:minimum_events_number_per_Xmax}. We assumed the aforementioned reconstruction resolutions.
\begin{table}[h!]
    \caption{Minimum number of events needed to distinguish between hadronic interaction models for different bins in \Xmax and two realistic mixed composition scenarios. \Xmax bins for which models are indistinguishable due to the model dependence of the bias in \LambdaMu are indicated by $-$.}\label{tab:minimum_events_number_per_Xmax}
    \centering
    \begin{tabular}{C{1.5cm}|C{1cm}C{1cm}|C{1cm}C{1cm}}
        & \multicolumn{2}{C{2cm}|}{$1:3:1:0$} & \multicolumn{2}{C{2cm}}{$7:1:2:0$} \\ \hline
        \Xmax \par $(\mathrm{g\,cm^{-2}})$ & $n_{\min}^{1 \sigma}$ & $n_{\min}^{3 \sigma}$ & $n_{\min}^{1 \sigma}$ & $n_{\min}^{3 \sigma}$  \\ \hline
        700 & $-$ & $-$ & $-$ & $-$ \\
        775 & $-$ & $-$ & $-$ & $-$ \\
        825 & 13\,030 & $100\,000$ & 18\,478 & $100\,000$ \\
        875 & 5\,080 & 54\,393 & 3\,519 & 29\,587 \\
        1100 & 3\,113 & 25\,898 & 1\,877 & 18\,805 \\
    \end{tabular}
\end{table}

\par
Since the models are close to indistinguishable for showers in the bulk of the \Xmax distribution, the number of events needed for a $3 \sigma$ model separation is greater than the number of events in the samples. However, about $3\,000$ events allow a $1 \sigma$ distinction between models in the deepest $X_{\max}$ bin, even in the proton-poor scenario. A $9$-fold increase in statistics allows for a $3 \sigma$ distinction. Note that the acceptance and quality cuts applied to data from different experiments might require more events than those estimated here. Nevertheless, these numbers of events are within the reach of the Pierre Auger Observatory \cite{ThePierreAuger:2020prl_sd1500spectrum, ThePierreAuger:2021_sd750spectrum}. For this experiment, the shower-to-shower estimations of $N_\mu$ and \Xmax could be provided by neural networks \cite{ThePierreAuger:2024prd_MassComp, ThePierreAuger:2024prl_MassInf, ThePierreAuger:2021jinst_xmax, ThePierreAuger:2021jinst_nmu}, shower universality \cite{Stadelmaier:icrc2023} and/or the improved detectors of AugerPrime~\cite{ThePierreAuger:2019_epj_augerprime}.

\section{Conclusions}

The variables describing the hadronic activity in first proton-air ($p$-air) interactions --- like multiplicity, inelasticity, and fraction of the primary energy going into hadronically interacting secondaries ---
significantly shape the two-dimensional distribution of the variables $X_{\max}$ and $N_{\mu}$ in ensembles of showers.

In this article, we show that the steepness of the distribution of the number of muons in muon-depleted showers, quantified by the parameter $\Lambda_{\mu}$, evolves with the depth of the shower maximum, \Xmax.
The different values of $\Lambda_\mu$ are due to the neutral pion energy spectrum emerging from different kinematic-phase space regions of the first $p$-air interaction.
Moreover, we verified that the value of $\Lambda_{\mu}$ for shallow showers is independent of the high-energy hadronic interaction model. On the other hand, for deep showers, $\Lambda_{\mu}$ shows an increased dependence on the specific interaction model used, reflecting the particular physical mechanisms it employs.

Lastly, we have established that the unbiased measurement of $\Lambda_{\mu}$ is achievable with current extensive air shower detectors, given their resolution, exposure and the mass interpretation of the cosmic ray flux. Therefore, the sensitivity of \LambdaMu to the hadronic activity of ultra-high-energy cosmic-ray air interactions, probed through \Xmax, establishes its potential to characterise hadronic interactions at energy scales beyond the reach of current human-made accelerators.


\section*{Acknowledgements}
We thank Carola Dobrigkeit, Gonzalo Parente, and Sofia Andringa for carefully reading this manuscript and their useful suggestions. We extended our gratitude to the Auger-IGFAE, Auger-LIP, and the Pierre Auger Collaboration members for their valuable insights throughout the different stages of this work. The authors thank Ministerio de Ciencia e Innovaci\'on/Agencia Estatal de Investigaci\'on
(PID2022-140510NB-I00 and RYC2019-027017-I), Xunta de Galicia (CIGUS Network of Research Centers,
Consolidaci\'on 2021 GRC GI-2033, ED431C-2021/22 and ED431F-2022/15),
and the European Union (ERDF).
This project was partly funded by Funda\c{c}\~ao para a Ci\^encia e Tecnologia under project 2024.06879.CERN.
MAM acknowledges that the project that gave rise to these results received the support of a fellowship from ``la Caixa” Foundation (ID 100010434). The fellowship code is LCF/BQ/DI21/11860033. FR received funding from the European Union’s Horizon 2020 research and innovation
programme under the Marie Skłodowska-Curie grant agreement No. 101065027.

\bibliography{paper}

\end{document}